\documentclass[twocolumn,aps,showpacs]{revtex4}

\usepackage{amsmath}
\usepackage{epsfig}
\usepackage{graphicx}

\begin{document}

\title{Three-point susceptibilities $\chi_n(k;t)$ and $\chi_n^s(k;t)$: 
mode-coupling approximation}
\author{Grzegorz Szamel and Elijah Flenner}
\affiliation{Department of Chemistry, 
Colorado State University, Fort Collins, CO 80523}

\date{\today}


\begin{abstract}
Recently, it was argued that a three-point susceptibility equal to the density
derivative of the intermediate scattering function, 
$\chi_n(k;t) = d F(k;t)/d n$, enters into an expression 
for the divergent part of an integrated four-point dynamic density
correlation function of a colloidal suspension
[Berthier \textit{et al.}, J. Chem. Phys. \textbf{126}, 184503 (2007)].
We show that, within the mode-coupling theory, 
the equation of motion for $\chi_n(k;t)$ is essentially identical as
the equation of motion for the $\mathbf{q}\to 0$ limit of the 
three-point susceptibility $\chi_{\mathbf{q}}(\mathbf{k};t)$
introduced by Biroli \textit{et al.} [Phys. Rev. Lett. \textbf{97},
195701 (2006)]. We present a numerical solution of 
the equation of motion for $\chi_n(k;t)$. 
We also derive and numerically solve an equation of motion for the density
derivative of the self-intermediate scattering function, 
$\chi_n^s(k;t) = d F^s(k;t)/d n$. We contrast the wave vector dependence 
of $\chi_n(k;t)$ and $\chi_n^s(k;t)$.
\end{abstract}
\maketitle

\section{Introduction}

While it has been generally accepted for quite some time that 
upon approaching the glass transition the liquid's dynamics 
become increasingly heterogeneous \cite{reviews},
theoretical descriptions of various four-point correlation 
functions which quantify dynamic heterogeneity have been scarce
\cite{GlotzerFranz,Franz}. It has been only in the last few years
that a number of different theoretical approaches to this problem appeared
\cite{BB, BerthierS, BBMR, IwataSasa, JCPa, JCPb, GSpairdiagram}.
According to one of these approaches \cite{JCPa, JCPb}, the divergent 
part of a four-point dynamic density correlation function, integrated over the whole
space, can be expressed
in terms of so-called three-point susceptibilities. These susceptibilities
are defined as the derivatives of the usual two-point correlation
function, \textit{i.e.} the intermediate scattering function, with
respect to the densities of the conserved quantities. In the case of a 
colloidal system there is only one conserved quantity, the number of particles and 
thus there is only one susceptibility, the density derivative
of the intermediate scattering function. 

The first goal of this note
is to derive the equation of motion for this susceptibility 
within the mode-coupling approximation \cite{Goetze,Shankar,SL} and to present
numerical solutions of this equation. Not surprisingly,
the equation of motion for the the density derivative
of the intermediate scattering function is 
essentially identical to that for the long-wave vector 
limit of the three-point susceptibility introduced by Biroli \textit{et al.} \cite{BBMR}. 
The latter susceptibility was defined as the derivative of the intermediate
scattering function with respect to an external, inhomogeneous potential.

It should be noted that while most theoretical approaches consider the 
full (\textit{i.e.} \textit{collective}) 
four-point dynamic density correlation function, 
in computer simulations usually the self part of this function is monitored
\cite{Glotzersc}. For this reason, while discussing 
the wave vector dependence of dynamic heterogeneities, Berthier \textit{et al.}
used a three-point
susceptibility equal to the density derivative of the self-intermediate 
scattering function (see Ref. \cite{JCPb}, Sec. IIC). 
Parenthetically, in this way they introduced a somewhat \textit{ad-hoc} 
extension of their theoretical approach. The approach of Berthier \textit{et al.},
at least as presented in Refs. \cite{JCPa, JCPb},
is only applicable to the full four-point function. 

The second goal of this note is to derive and numerically solve 
the equation of motion for
the density derivative of the self-intermediate scattering function
within the mode-coupling approximation, and to contrast the time and 
wave vector dependence of the density derivatives of the collective and
self-intermediate scattering functions. In Ref. \cite{JCPb}
the density derivative of the self-intermediate scattering function 
was calculated by solving the mode-coupling equation
of motion for this function at different densities and then performing 
a numerical differentiation. Numerical solution of the equation of motion for 
the density derivative of the self-intermediate scattering function 
allows us to investigate 
the wave vector at which, according to 
Berthier \textit{et al.} \cite{JCPa, JCPb}, the dynamics of a glassy colloidal
suspension is maximally heterogeneous. 

\section{Equations of motion for $\chi_n(k;t)$ and $\chi_n^s(k;t)$}

We consider the simplest possible model of a colloidal suspension: 
a system of interacting Brownian particles. 
Our first object of interest is a three-point susceptibility $\chi_n(k;t)$ which is 
defined as the density derivative of the intermediate scattering function,
\begin{equation}\label{defchin}
\chi_n(k;t) = \frac{d F(k;t)}{d n}.
\end{equation}
Here $F(k;t)$ denotes the intermediate scattering function,
\begin{equation}\label{defF}
F(k;t) = \frac{1}{N} \left<n(\mathbf{k};t) n(-\mathbf{k})\right>
\end{equation}
where $N$ is the number of particles,
$n(\mathbf{k};t)$ is the Fourier transform of the microscopic
density at time $t$ and $n(\mathbf{k}) \equiv n(\mathbf{k};t=0)$.
Finally, in Eq. (\ref{defchin}) $n$ is the number density, $n=N/V$, with 
$V$ being the volume of the system (the particle diameter $\sigma$ is used
as the unit of length and thermodynamic limit is implied throughout).
The initial value of the intermediate scattering function
is equal to the static structure factor $S(k)$, $F(k;t=0) = S(k)$.

Using standard, exact but formal, methods (\textit{e.g.} a projection
operator approach) one can derive the following equation of motion for 
the intermediate scattering function,
\begin{eqnarray}\label{eomF}
&& 
\int_0^t dt' (\delta(t-t') + M^{\mathrm{irr}}(k;t-t')) \partial_{t'} F(k;t') 
=  \nonumber \\ &&
- \frac{ D_0 k^2}{S(k)} F(k;t).
\end{eqnarray}
where $D_0$ is the diffusion coefficient of an isolated Brownian particle and 
$M^{\mathrm{irr}}(k;t)$ is the so-called irreducible \cite{irr} memory function.
Within the mode-coupling approximation \cite{Goetze,Shankar} applied to the
the system of interacting Brownian particles, the irreducible memory function is 
given by the following expression \cite{SL}
\begin{eqnarray}\label{MCTM}
\lefteqn{M^{\mathrm{irr}}(k;t) = } \nonumber \\ &&  
\frac{n D_0 }{2} \int \frac{d\mathbf{k}_1}{(2\pi)^3} 
\left(c(k_1)\hat{\mathbf{k}}\cdot\mathbf{k}_1+
c(|\mathbf{k}-\mathbf{k}_1|)\hat{\mathbf{k}}\cdot(\mathbf{k}-\mathbf{k}_1)\right)^2 
\nonumber \\ && \times
F(k_1;t) F(|\mathbf{k}-\mathbf{k}_1|;t).
\end{eqnarray}
Here $\hat{\mathbf{k}} = \mathbf{k}/k$ and 
$c(k)$ is the so-called direct correlation function,
$c(k) = (1-1/S(k))/n$. 

Substituting Eq. (\ref{MCTM}) into Eq. (\ref{eomF}) and then differentiating
the resulting equation with respect to the density we obtain the following 
equation of motion for the three-point susceptibility $\chi_n(k;t)$,
\begin{widetext}
\begin{eqnarray}\label{eomchin}
\lefteqn{
\int_0^t dt' (\delta(t-t') + M^{\mathrm{irr}}(k;t-t')) \partial_{t'} \chi_n(k;t')
+ \frac{ D_0 k^2}{S(k)} \chi_n(k;t) }  \nonumber \\ && + 
n D_0 \int_0^t \int \frac{d\mathbf{k}_1}{(2\pi)^3}
\left(c(k_1)\hat{\mathbf{k}}\cdot\mathbf{k}_1+
c(|\mathbf{k}-\mathbf{k}_1|)\hat{\mathbf{k}}\cdot(\mathbf{k}-\mathbf{k}_1)\right)^2 
\chi_n(k_1;t-t') F(|\mathbf{k}-\mathbf{k}_1|;t-t')\partial_{t'} F(k;t')
\nonumber \\ &=& 
\frac{ D_0 k^2}{S^2(k)} \frac{d S(k) }{dn} F(k;t) \nonumber \\ && 
- \frac{D_0}{2} \int_0^t dt' \int \frac{d\mathbf{k}_1}{(2\pi)^3} 
\left(c(k_1)\hat{\mathbf{k}}\cdot\mathbf{k}_1+
c(|\mathbf{k}-\mathbf{k}_1|)\hat{\mathbf{k}}\cdot(\mathbf{k}-\mathbf{k}_1)\right)^2 
F(k_1;t-t') F(|\mathbf{k}-\mathbf{k}_1|;t-t') \partial_{t'} F(k;t')
\nonumber \\ && -
nD_0 \int_0^t dt' \int \frac{d\mathbf{k}_1}{(2\pi)^3} 
\left(c(k_1)\hat{\mathbf{k}}\cdot\mathbf{k}_1+
c(|\mathbf{k}-\mathbf{k}_1|)\hat{\mathbf{k}}\cdot(\mathbf{k}-\mathbf{k}_1)\right)
\left(\frac{d c(k_1)}{dn} \hat{\mathbf{k}}\cdot\mathbf{k}_1+
\frac{d c(|\mathbf{k}-\mathbf{k}_1|)}{dn}
\hat{\mathbf{k}}\cdot(\mathbf{k}-\mathbf{k}_1)\right) \nonumber \\ && \times
F(k_1;t-t') F(|\mathbf{k}-\mathbf{k}_1|;t-t')\partial_{t'} F(k;t')
\end{eqnarray}
\end{widetext}

We notice that the left-hand-side of Eq. (\ref{eomchin}) is exactly the same
as the left-hand-side of 
the overdamped limit of the equation of motion derived by Biroli \textit{et al.}
\cite{BBMR} for their three-point susceptibility $\chi_{\mathbf{q}}(\mathbf{k};t)$
in the $\mathbf{q}\to 0$ limit. The three-point susceptibility 
$\chi_{\mathbf{q}}(\mathbf{k};t)$ is defined as the derivative of the 
intermediate scattering function with respect to a static, inhomogeneous
external potential. 

Furthermore, we recall that the derivation of the mode-coupling expression
for the irreducible memory function invokes the convolution approximation
for the three-point equilibrium density correlation function \cite{SL}. 
The convolution approximation is equivalent to assuming that the 
direct correlation function is density-independent \cite{BHP}. 
To be consistent, we have to use the same approximation in Eq. (\ref{eomchin}).
Using $d c(k)/dn =0$ and $d S(k)/dn \equiv S^2(k) (c(k) + n d c(k)/dn) = S^2(k) c(k)$ 
in Eq. (\ref{eomchin}) we obtain the 
following equation of motion for the three-point susceptibility $\chi_n(k;t)$
\begin{widetext}
\begin{eqnarray}\label{eomchins}
\lefteqn{
\int_0^t dt' (\delta(t-t') + M^{\mathrm{irr}}(k;t-t')) \partial_{t'} \chi_n(k;t')
+ \frac{ D_0 k^2}{S(k)} \chi_n(k;t) }  \nonumber \\ && + 
n D_0 \int_0^t \int \frac{d\mathbf{k}_1}{(2\pi)^3}
\left(c(k_1)\hat{\mathbf{k}}\cdot\mathbf{k}_1+
c(|\mathbf{k}-\mathbf{k}_1|)\hat{\mathbf{k}}\cdot(\mathbf{k}-\mathbf{k}_1)\right)^2 
\chi_n(k_1;t-t') F(|\mathbf{k}-\mathbf{k}_1|;t-t')\partial_{t'} F(k;t')
\nonumber \\ &=& 
D_0 k^2 c(k) F(k;t) \nonumber \\ && 
- \frac{D_0}{2} \int_0^t dt' \int \frac{d\mathbf{k}_1}{(2\pi)^3} 
\left(c(k_1)\hat{\mathbf{k}}\cdot\mathbf{k}_1+
c(|\mathbf{k}-\mathbf{k}_1|)\hat{\mathbf{k}}\cdot(\mathbf{k}-\mathbf{k}_1)\right)^2 
F(k_1;t-t') F(|\mathbf{k}-\mathbf{k}_1|;t-t') \partial_{t'} F(k;t')
\end{eqnarray}
\end{widetext}

At this point we notice that the difference between 
the right-hand-side of Eq. (\ref{eomchins}) and the right-hand-side of 
the equation of motion derived by Biroli \textit{et al.}
\cite{BBMR} for the three-point susceptibility 
$\chi_{\mathbf{q}}(\mathbf{k};t)$ in the $\mathbf{q}\to 0$ limit is
a constant factor equal to $-nS(0)$. This factor is equal to 
the thermodynamic derivative, $-nS(0) = - \left(\partial n /\partial \beta \mu\right)_T$
where $\beta = 1/(k_B T)$ and $\mu$ is the chemical potential. 
To rationalize this fact we 
notice that Biroli \textit{et al.} used approximations which in the 
limit of $\mathbf{q}\to 0$ amount to $d c(k)/dn =0$. Furthermore, their 
three-point susceptibility is defined as the derivative of the intermediate 
scattering function with respect to the external potential. 
In the long wavelength limit, $\mathbf{q}\to 0$, this derivative 
differs from the derivative
with respect to the density by a thermodynamic factor proportional to 
$\left(\partial n /\partial \beta \mu\right)_T$. 

Our second object of interest is a three-point susceptibility 
$\chi_n^s(k;t)$ which is 
defined as the density derivative of the self-intermediate scattering function,
\begin{equation}\label{defchins}
\chi_n^s(k;t) = \frac{d F^s(k;t)}{d n}.
\end{equation}
Here $F^s(k;t)$ denotes the self-intermediate scattering function,
\begin{equation}\label{defFs}
F^s(k;t) = \left<n_1(\mathbf{k};t) n_1(-\mathbf{k})\right>,
\end{equation}
where $n_1(\mathbf{k};t)$ is the Fourier transform of the microscopic
density of one selected (labeled) particle at time $t$ 
and $n_1(\mathbf{k}) \equiv n_1(\mathbf{k};t=0)$. The initial value 
of the self-intermediate scattering function is equal to 1, $F^s(k;t=0)=1$.

The derivation of the equation of motion for $\chi_n^s(k;t)$ is analogous 
to that for $\chi_n(k;t)$. Here we only present the starting point, \textit{i.e.}
the equation of motion for $F^s(k;t)$ and 
the final result, \textit{i.e.} the equation of motion for $\chi_n^s(k;t)$.
The equation of motion for the self-intermediate scattering function
$F^s(k;t)$ reads
\begin{eqnarray}\label{eomFs}
&& 
\int_0^t dt' (\delta(t-t') + M^{s\mathrm{irr}}(k;t-t')) \partial_{t'} F^s(k;t') 
=  \nonumber \\ &&
- D_0 k^2 F^s(k;t),
\end{eqnarray}
where the self irreducible memory function is given by the following expression,
\begin{eqnarray}\label{sMCTM}
\lefteqn{M^{s\mathrm{irr}}(k;t) = }  \\ \nonumber &&  
n D_0  \int \frac{d\mathbf{k}_1}{(2\pi)^3} 
\left(c(k_1)\hat{\mathbf{k}}\cdot\mathbf{k}_1\right)^2 
F(k_1;t) F^s(|\mathbf{k}-\mathbf{k}_1|;t).
\end{eqnarray}
The equation of motion for three-point susceptibility $\chi_n^s(k;t)$ has the following
form,
\begin{widetext}
\begin{eqnarray}\label{eomchinselfs}
\lefteqn{
\int_0^t dt' (\delta(t-t') + M^{s\mathrm{irr}}(k;t-t')) \partial_{t'} \chi_n^s(k;t')
+ D_0 k^2 \chi_n^s(k;t) }  \nonumber \\ && + 
n D_0 \int_0^t \int \frac{d\mathbf{k}_1}{(2\pi)^3}
\left(c(k_1)\hat{\mathbf{k}}\cdot\mathbf{k}_1\right)^2 
F(k_1;t-t') \chi_n^s(|\mathbf{k}-\mathbf{k}_1|;t-t')\partial_{t'} F^s(k;t')
\nonumber \\ &=& 
- D_0 \int_0^t dt' \int \frac{d\mathbf{k}_1}{(2\pi)^3} 
\left(c(k_1)\hat{\mathbf{k}}\cdot\mathbf{k}_1\right)^2 
F(k_1;t-t') F^s(|\mathbf{k}-\mathbf{k}_1|;t-t') \partial_{t'} F^s(k;t')
\nonumber \\ && -
nD_0 \int_0^t dt' \int \frac{d\mathbf{k}_1}{(2\pi)^3} 
\left(c(k_1)\hat{\mathbf{k}}\cdot\mathbf{k}_1 \right)^2
\chi_n(k_1;t-t') F^s(|\mathbf{k}-\mathbf{k}_1|;t-t')\partial_{t'} F^s(k;t')
\end{eqnarray}
\end{widetext}
Note that to derive Eq. (\ref{eomchinselfs}) we again used $dc(k)/dn=0$.

\section{Numerical solutions of equations of motion for $\chi_n(k;t)$
and $\chi_n^s(k;t)$}

Equations of motion (\ref{eomchins}) and (\ref{eomchinselfs}) 
for the three-point susceptibilities 
$\chi_n(k;t)$ and $\chi_n^s (k;t)$ can be solved using the algorithm used previously 
\cite{FuchsMCT,Miyazaki,FlennerSzamel} to solve mode-coupling 
equations (\ref{eomF}-\ref{MCTM}) for the intermediate scattering function. 
The only input required is the structure factor $S(k)$. Here, in order to be consistent
with earlier related work \cite{JCPb}, we use the structure factor calculated
for the hard sphere interaction potential 
using the Percus-Yevick closure approximation. As is customary for 
a mode-coupling calculation for the hard sphere system, we report the results using 
volume fraction $\phi = n\pi \sigma^3/6$ where $\sigma$ is the hard sphere
diameter or relative distance from the mode-coupling transition 
$\epsilon = (\phi_c-\phi)/\phi_c$.

To solve equations of motion (\ref{eomchins}) and (\ref{eomchinselfs}) we used
300 equally spaced wave vectors from $k=0$ to $k=60$
with the first wave vector $k_0 = 0.1$.
For this discretization of the mode-coupling equations
(\ref{eomF}-\ref{MCTM}) the ergodicity breaking transition (\textit{i.e.} the 
mode-coupling transiton) is
located at volume fraction $\phi_c = 0.515866763$. We also 
performed a few calculations with
larger cutoffs for the integral and/or a finer grid of
wave vectors and obtained qualitatively the same results.

\begin{figure}
\includegraphics[scale=0.8]{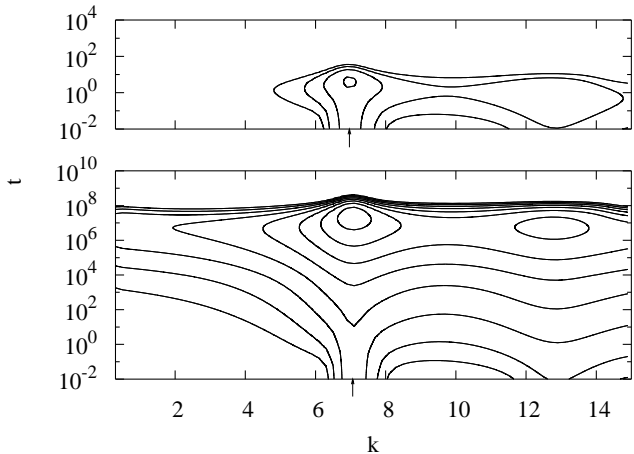}
\caption{\label{chintk} Three-point susceptibility $\chi_n(k;t)$ for 
$\epsilon = 0.05$ (upper panel) and $\epsilon = 10^{-4}$ (lower panel). 
Contours correspond to $\chi_n(k;t)= 4^m$ where $m$ is an integer, starting from $m=-1$.
The arrow marks the position of the first peak of the 
static structure factor.}
\end{figure}

\begin{figure}
\includegraphics[scale=0.8]{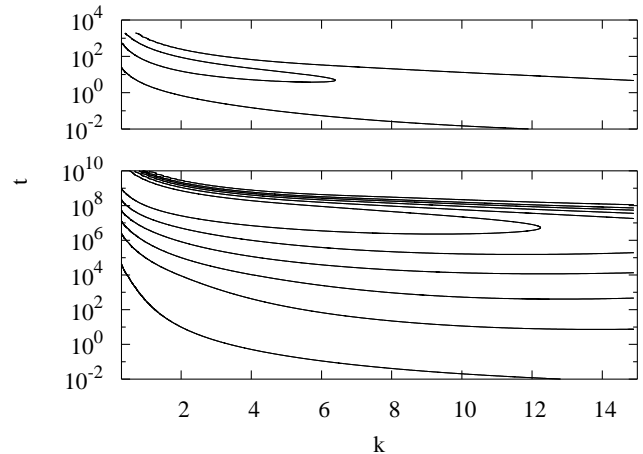}
\caption{\label{chintks} Three-point susceptibility $\chi_n^s(k;t)$ for 
$\epsilon = 0.05$ (upper panel) and $\epsilon = 10^{-4}$ (lower panel). 
Contours correspond to
$\chi_n^s(k;t)= 4^m$ where $m$ is an integer, starting from $m=-1$. }
\end{figure}

In Figs. \ref{chintk} and \ref{chintks} we show $\chi_n(k;t)$ and $\chi_n^s(k;t)$,
respectively, as functions of both time $t$ and 
wave vector $k$ for two different values of $\epsilon = (\phi_c - \phi)/\phi_c$,
$\epsilon = 0.05$ and $\epsilon = 10^{-4}$. The former value of $\epsilon$ 
is comparable to the lowest reduced temperature $(T-T_c)/T_c$ at which 
mode-coupling theory agrees \cite{FlennerSzamel} 
with computer simulations for the well-known
glass former, the Kob-Andersen Lennard-Jones binary mixture \cite{KobAndersen}. 
Using contour plots to show these three-point susceptibilities 
is inspired by a similar presentation used
by Lechenault \textit{et al.}\ \cite{LDBB} in their investigation 
of  dynamic heterogeneity in dense, two-dimensional granular systems.
One should note, however, that Lechenault \textit{et al.} monitored and  
analyzed a correlation function of the fluctuations of the self-overlap 
integrated over the whole space. Here, we present results for 
three-point susceptibilities and we need to invoke the theory of Berthier
\textit{et al.}\ to connect these susceptibilities to dynamic heterogeneity.

We note that, as expected, the susceptibility $\chi_n(k;t)$ has a well-defined
maximum at a characteristic time and wave vector. The characteristic
time strongly increases upon approaching the mode-coupling transition whereas
the characteristic wave vector is approximately independent of the distance
from the transition and equal to 7.1 close to the transition (note that according to the 
Percus-Yevick approximation the first peak of the 
static structure factor at $\phi_c$ is located at $k= 7.1$).

Examining Fig. \ref{chintks} we note that, somewhat surprisingly,
$\chi_n^s(k;t)$ does not have a well-defined maximum at a characteristic time and 
wave vector. In contrast with $\chi_n(k;t)$, $\chi_n^s(k;t)$
as a function of time and wave vector forms a ridge which, with decreasing wave vector,
gently increases and moves towards longer times. For a fixed wave vector,
\textit{e.g.} for $k=k_{\mathrm{max}}= 7.1$ 
(\textit{i.e.} the position of the first peak of 
the structure factor at $\phi_c$), the peak position 
of $\chi_n^s(k;t)$ strongly increases upon approaching the mode-coupling transition.

\begin{figure}
\includegraphics[scale=0.3]{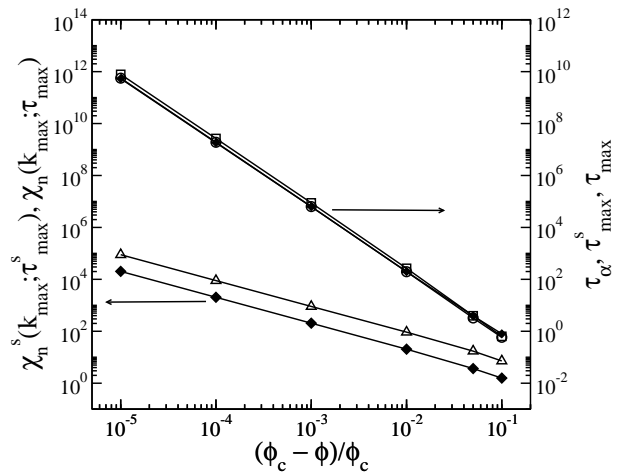}
\caption{\label{tkmaxeps} Right scale: dependence of the peak time of  
$\chi_n(k_{\mathrm{max}};t)$ and $\chi_n^s(k_{\mathrm{max}};t)$ , 
$\tau_{\mathrm{max}}$ and $\tau_{\mathrm{max}}^s$, respectively, and of the 
$\alpha$ relaxation time \cite{talpha}, $\tau_{\alpha}$, on the reduced 
distance from the mode-coupling transition, $\epsilon = (\phi_c-\phi)/\phi_c$.
Squares: $\tau_{\mathrm{max}}$; filled diamonds: $\tau_{\mathrm{max}}^s$;
circles: $\tau_{\alpha}$. Left scale: dependence of the peak value
of three-point susceptibilities, $\chi_n(k_{\mathrm{max}}; \tau_{\mathrm{max}})$
and $\chi_n(k_{\mathrm{max}}; \tau_{\mathrm{max}})$
on $\epsilon$. Open triangles: $\chi_n(k_{\mathrm{max}}; \tau_{\mathrm{max}})$; 
filled diamonds: $\chi_n^s(k_{\mathrm{max}}; \tau_{\mathrm{max}}^s)$.}
\end{figure}

In Fig. \ref{tkmaxeps} we analyze in some detail 
the dependence of both three-point susceptibilities on the reduced distance from the
mode-coupling transition for a fixed wave vector equal to the position of 
the first peak of the static structure at the transition, $k=k_{\mathrm{max}} = 7.1$.
We show the dependence on $\epsilon = (\phi_c-\phi)/\phi_c$ of the time at which  
$\chi_n(k_{\mathrm{max}};t)$ and $\chi_n^s(k_{\mathrm{max}};t)$ has the maximum
value, $\tau_{\mathrm{max}}$
and $\tau_{\mathrm{max}}^s$, respectively, and we compare these times to the 
$\alpha$ relaxation time \cite{talpha}, $\tau_{\alpha}$. 
As expected, both $\tau_{\mathrm{max}}$ and $\tau_{\mathrm{max}}^s$
have the same $\epsilon$ dependence as  $\tau_{\alpha}$, 
$\tau_{\mathrm{max}} \sim \tau_{\mathrm{max}}^s \sim \tau_{\alpha} \sim \epsilon^{-2.5}$ 
and ratios of these times and $\tau_{\alpha}$ are approximately constant, 
$\tau_{\mathrm{max}}/\tau_{\alpha} \approx 1.4$ and 
$\tau_{\mathrm{max}}^s/\tau_{\alpha} \approx 0.96$.
In addition, we show in Fig. \ref{tkmaxeps} the dependence on $\epsilon$ 
of the peak value
of the three-point susceptibilities for a fixed wave vector $k=k_{\mathrm{max}} = 7.1$. 
As expected, we find 
$\chi_n(k_{\mathrm{max}}; \tau_{\mathrm{max}}) 
\sim \chi_n^s(k_{\mathrm{max}}; \tau_{\mathrm{max}}^s) \sim\epsilon^{-1}$. 

\begin{figure}
\includegraphics[scale=0.3]{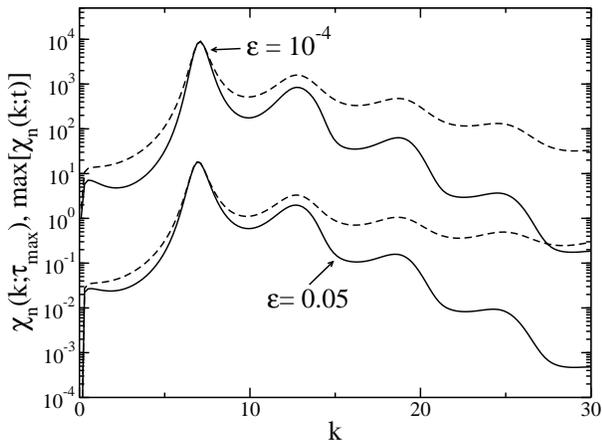}
\caption{\label{chitmaxk} Wave vector dependence of the 
$\chi_n(k;\tau_{\mathrm{max}})$ (solid curves) and of the maximum value of 
$\chi_n(k;t)$ (dashed curves) for $\epsilon = 0.05$ (lower curves) 
and $\epsilon = 10^{-4}$ (upper curves). }
\end{figure}

\begin{figure}
\includegraphics[scale=0.3]{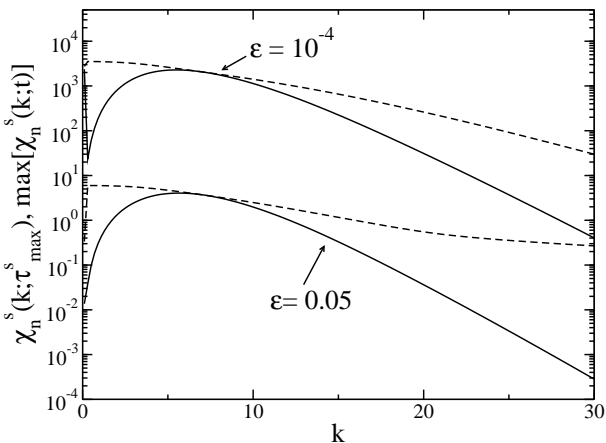}
\caption{\label{chitmaxks} Wave vector dependence of the 
$\chi_n^s(k;\tau_{\mathrm{max}}^s)$ (solid curves) and of the maximum value of 
$\chi_n^s(k;t)$ (dashed curves) for $\epsilon = 0.05$ (lower curves) 
and $\epsilon = 10^{-4}$ (upper curves). }
\end{figure}

The original motivation for investigating two-dimensional (contour) plots of 
$\chi_n(k;t)$ and $\chi_n^s(k;t)$ was to find the characteristic wave vector
at which these susceptibilities have maximum values. Once these wave vectors
have been found, one could use the theory of Berthier \textit{et al.} to 
claim that at these wave vectors dynamic heterogeneities are the strongest. 
From this perspective, Figs. \ref{chintk}-\ref{chintks} lead to a rather 
surprising conclusion: at a fixed time comparable to the $\alpha$ relaxation 
time both $\chi_n(k;t)$ and $\chi_n^s(k;t)$ have a maximum at a nonzero 
wave vector. However, Fig. \ref{chintks} suggests that the absolute
maximum of $\chi_n^s(k;t)$ is located at or close to $k=0$. The difference between
the wave vector dependence of $\chi_n(k;t)$ and $\chi_n^s(k;t)$ is investigated in
some detail in Figs. \ref{chitmaxk} and \ref{chitmaxks}. In the former figure
we compare the wave vector dependence of $\chi_n(k;\tau_{\mathrm{max}})$
(recall that $\tau_{\mathrm{max}}$ is the peak position of $\chi_n(k_{\mathrm{max}};t)$)
with the wave vector dependence of the maximum value of $\chi_n(k;t)$
for two different values of $\epsilon=(\phi_c-\phi)/\phi_c$, $\epsilon=0.05$
and $\epsilon=10^{-4}$. We notice that the wave vector dependence is quite 
strong but qualitatively similar, except for higher wave vectors. 
In Fig. \ref{chitmaxks} we show the wave vector dependence of 
$\chi_n^s(k;\tau_{\mathrm{max}})$  ($\tau_{\mathrm{max}}^s$ is the peak position 
of $\chi_n^s(k_{\mathrm{max}};t)$) with the wave vector dependence of the maximum 
value of $\chi_n^s(k;t)$ for the same two different values of $\epsilon$.
We see that in this case the wave vector dependence is quite a bit weaker
than that shown in Fig. \ref{chitmaxk}. More importantly, we see that the 
wave vector dependence of $\chi_n^s(k;\tau_{\mathrm{max}})$ and 
of the the maximum value of $\chi_n^s(k;t)$ are qualitatively different.

\begin{figure}
\includegraphics[scale=0.3]{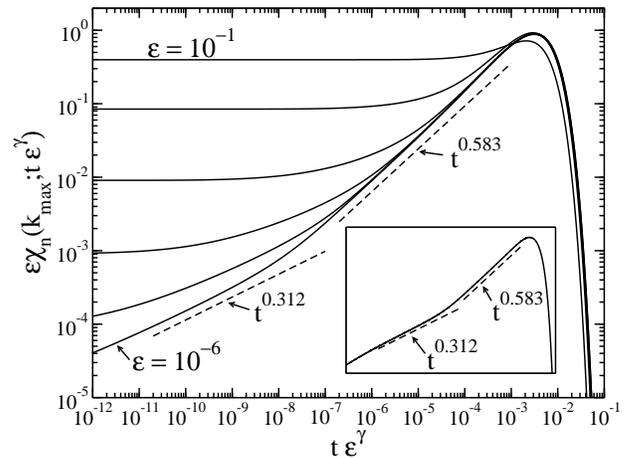}
\caption{\label{chiscale} Scaling plot of $\chi_n(k_{\mathrm{max}};t)$. 
Solid lines correspond to $\epsilon$ equal to 
$10^{-1}, 10^{-2}, 10^{-3}, 10^{-4}, 10^{-5}$, and $10^{-6}$
(top to bottom). 
Inset: time dependence of $\chi_n^s(k_{\mathrm{max}};t)$ at a very small
value of $\epsilon = 10^{-6}$. Dashed lines 
in the main figure and in the inset indicate 
transient power law dependence in the early and late $\beta$ regimes. }
\end{figure}

Finally, in Fig. \ref{chiscale}
we show that upon approaching the mode-coupling transition 
three-point susceptibility $\chi_n(k_{\mathrm{max}};t)$ 
approaches the scaling limit. Three point susceptibility $\chi_n^s(k_{\mathrm{max}};t)$
also approaches the scaling limit, but in a somewhat more complicated way 
(not shown). The inset in Fig. \ref{chiscale} shows the time dependence 
of $\chi_n^s(k_{\mathrm{max}};t)$ for a small
value of $\epsilon$ equal to $10^{-6}$. Comparison of the main figure and the inset 
shows that upon approaching the mode-coupling transition 
the time dependence of both susceptibilities at $k=k_{\mathrm{max}}$ 
is quite similar.

\section{Discussion} 

The main reason for our interest in three-point susceptibilities 
$\chi_n(k;t)$ and $\chi_n^s(k;t)$ was the connection between
these susceptibilities and  
four-point correlation functions which quantify dynamic heterogeneities.
Roughly speaking, according to Berthier \textit{et al.} \cite{JCPa,JCPb}, 
Brownian system's 
four-point correlation function $\chi_4(k;t)$, which is defined as the
self part of the four-point dynamic density correlation function integrated
over the whole space, 
is proportional to $\chi_n^s(k;t)^2$ for a system in which the
total number of particles can fluctuate and is proportional to $\chi_n^s(k;t)$
for a system with fixed total number of particles (note that the latter one 
is commonly used in numerical simulations). Following Berthier \textit{et al.}
and assuming that mode-coupling theory gives at least qualitatively correct
results for three-point susceptibility $\chi_n^s(k;t)$ we are faced with
a striking conclusion that the there is no finite \textit{intrinsic} \cite{CR}
wave vector at which dynamic heterogeneities, as quantified by $\chi_4(k;t)$, 
are the largest. More precisely,
at a fixed time we can determine a finite characteristic wave vector, 
but with increasing time this wave vector is decreasing towards 0. In other words, 
the maximum value of $\chi_4(k;t)$ is predicted to monotonically decrease
with increasing $k$.

The wave vector dependence of four-point correlation function $\chi_4(k;t)$
was monitored in two recent simulational investigations \cite{CR,CGJMP}. Both studies
used Newtonian rather than Brownian dynamics and thus our findings cannot
be directly compared to their results. However, it is interesting to note that
Ref. \cite{CGJMP} found that the time at which $\chi_4(k;t)$ reaches its
maximum value increases as the wave vector decreases from the peak position
of the structure factor, in rough agreement with Fig. \ref{chintks}.
The authors of Ref. \cite{CGJMP} did not comment on the dependence of 
the maximum value of $\chi_4(k;t)$ on the wave vector but from their Fig. 1
it seems that it does not increase with decreasing wave vector, in 
contrast to our Figs. \ref{chintks} and \ref{chitmaxks}. 
The second study, Ref. \cite{CR},
shows the dependence of the maximum value of $\chi_4(k;t)$ \cite{commentCR} 
on the wave vector. It exhibits a broad maximum located 
at a wave vector slightly smaller than the peak position of the 
static structure factor. Qualitatively, this disagrees with our results shown
in Figs. \ref{chintks} and \ref{chitmaxks}. At present, the origin of the 
differences between our findings and the results of Refs. \cite{CR,CGJMP} is unclear.

In closing, we would like to emphasize that 
various four-point functions quantifying dynamic heterogeneities have
been studied in computer simulations for more than a decade. On the 
other hand, theoretical predictions for these functions started to 
appear only in the last few years. Quantitative comparison between simulations
and theories is still in infancy. We hope that this work will stimulate 
further effort in this direction.  

\section*{Acknowledgments} 
We gratefully acknowledge the support of NSF Grant No.~CHE 0517709.


\begin{thebibliography}{99}
\bibitem{talpha} The $\alpha$ relaxation time, $\tau_{\alpha}$, 
is defined as the time at which
the intermediate scattering function $F(k;t)$ at the wave vector 
corresponding to the first peak of the static structure factor 
is equal to $e^{-1}$, $F(k_{\mathrm{max}};\tau_{\alpha}) = e^{-1}$.
\bibitem{commentCR} Note that the definition of $\chi_4(k;t)$ used in Ref. \cite{CR} 
is somewhat non-standard.
\end{thebibliography}
\end{document}